\documentclass[12pt,a4paper]{article}
\usepackage{blkarray}
\usepackage{amsfonts}
\usepackage{amsmath}
\usepackage{geometry}
\usepackage{graphicx}
\usepackage{amssymb}
\usepackage{abstract}
\usepackage{setspace}
\usepackage{geometry}
\providecommand{\U}[1]{\protect\rule{.1in}{.1in}}
\newtheorem{theorem}{Theorem}

\newtheorem{notation}[theorem]{Remark}

\def\fr{\frac}
\def\be{\begin{equation}}
\def\ee{\end{equation}}
\def\ba{\begin{eqnarray}}
\def\ea{\end{eqnarray}}
\def\pa{\partial}
\def\ra{\rightarrow}
\def\s{\sigma}
\def\l{\lambda}

\def\a{\alpha}
\def\b{\beta}
\def\g{\gamma}
\def\d{\delta}
\def\t{\tau}

\def\pt{\phantom{a}}
\def\w{\omega}
\def\L{\Lambda^4}

\def\f{\varphi}

\begin{document}

\title{Quantizable non-local gravity}

\author{Diego Marin \thanks{dmarin.math@gmail.com}}

\maketitle

\begin{abstract}\begin{spacing}{1.2}
It's widely recognized that general relativity emerges if we impose invariance under
local translations and local Lorentz transformations. In the same manner supergravity
arises when we impose invariance under local supersymmetry.
In this paper we show how to treat general relativity as a common gauge theory,
without introducing a metric or a tetrad field. The price to pay for such
simplification is the acceptance of non-locality. At first glance the resulting theory seems renormalizable.
Finally we derive Feynman vertices for such theory.
\end{spacing}
\end{abstract}

\newpage
\tableofcontents
\newpage

\section{Introduction}

It's widely recognized that general relativity emerges if we impose
invariance under local translations and local Lorentz transformations.
In the same manner supergravity arises when we impose invariance under
local supersymmetry.

In this paper we show how to treat general relativity as a common gauge theory,
without introducing a metric or a tetrad field. The price to pay for such
simplification is the acceptance of non-locality. Several indications in such
direction are coming from Arrangement Field Theory\cite{Arrangement}.

\section{Non local operators \label{nonlocal}}

In this section we define non-local fields over discrete spacetimes,
showing that such fields include differential operators.

Every euclidean $4$-dimensional space can be approximated by a graph $\L$, that
is a collection of vertices connected by edges of length $\Delta$. We recover the
continuous space in the limit $\Delta \ra 0$. Moreover we can pass from the euclidean
space to the lorenzian space-time by extending holomorphically any function in the
fourth coordinate $x_4 \ra ix_4$ \cite{minko}.

We start with a local scalar field $\varphi(p_i)$ represented
by a column array where each entry is the value of the field in a
specific vertex of the graph. For example (with only 5 vertices):

\begin{equation}
\varphi(p_i)=\left(
\begin{array}
[c]{c}%
\varphi\left(  p_{0}\right) \\
\varphi\left(  p_{1}\right) \\
\varphi\left(  p_{2}\right) \\
\varphi\left(  p_{3}\right) \\
\varphi\left(  p_{4}\right) \\
\end{array}
\right) \qquad \L = \{ p_0, p_1, p_2, p_3, p_4 \}
\end{equation}

\noindent Similarly we can define a non-local scalar field as:

\begin{equation}
\varphi (p_i,p_j)=\left(
\begin{array}
[c]{ccccc}
\varphi(p_{0},p_0)& \varphi(p_0,p_1)& \varphi(p_0,p_2)& \varphi(p_0,p_3)& \varphi(p_0,p_4) \\
\varphi(p_{1},p_0)& \varphi(p_1,p_1)& \varphi(p_1,p_2)& \varphi(p_1,p_3)& \varphi(p_1,p_4) \\
\varphi(p_{2},p_0)& \varphi(p_2,p_1)& \varphi(p_2,p_2)& \varphi(p_2,p_3)& \varphi(p_2,p_4) \\
\varphi(p_{3},p_0)& \varphi(p_3,p_1)& \varphi(p_3,p_2)& \varphi(p_3,p_3)& \varphi(p_3,p_4) \\
\varphi(p_{4},p_0)& \varphi(p_4,p_1)& \varphi(p_4,p_2)& \varphi(p_4,p_3)& \varphi(p_4,p_4) \\
\end{array}
\right)
\end{equation}

\noindent At the same time, a local field can be represented also by a diagonal matrix:

\begin{equation}
\varphi^D (p_i) = \varphi (p_i,p_i)=\left(
\begin{array}
[c]{ccccc}
\varphi(p_{0},p_0)& 0& 0& 0& 0 \\
0& \varphi(p_1,p_1)& 0& 0& 0 \\
0& 0& \varphi(p_2,p_2)& 0& 0 \\
0& 0& 0& \varphi(p_3,p_3)& 0 \\
0& 0& 0& 0& \varphi(p_4,p_4) \\
\end{array}
\right)
\end{equation}

\ba
\varphi (p_i) &=& \varphi^D (p_i) \cdot \vec{\mathbf{1}} = \nonumber \\
 &=& \left(
\begin{array}
[c]{ccccc}
\varphi(p_{0},p_0)& 0& 0& 0& 0 \\
0& \varphi(p_1,p_1)& 0& 0& 0 \\
0& 0& \varphi(p_2,p_2)& 0& 0 \\
0& 0& 0& \varphi(p_3,p_3)& 0 \\
0& 0& 0& 0& \varphi(p_4,p_4) \\
\end{array}
\right) \left(
\begin{array}
[c]{c}
1 \\
1 \\
1 \\
1 \\
1 \\
\end{array}
\right) \nonumber \\
&=& \left(
\begin{array}
[c]{c}
\varphi\left(  p_{0}\right) \\
\varphi\left(  p_{1}\right) \\
\varphi\left(  p_{2}\right) \\
\varphi\left(  p_{3}\right) \\
\varphi\left(  p_{4}\right) \\
\end{array}
\right)
\ea

\begin{figure}[ptbh]
\centering\includegraphics[width=0.4\textwidth ]{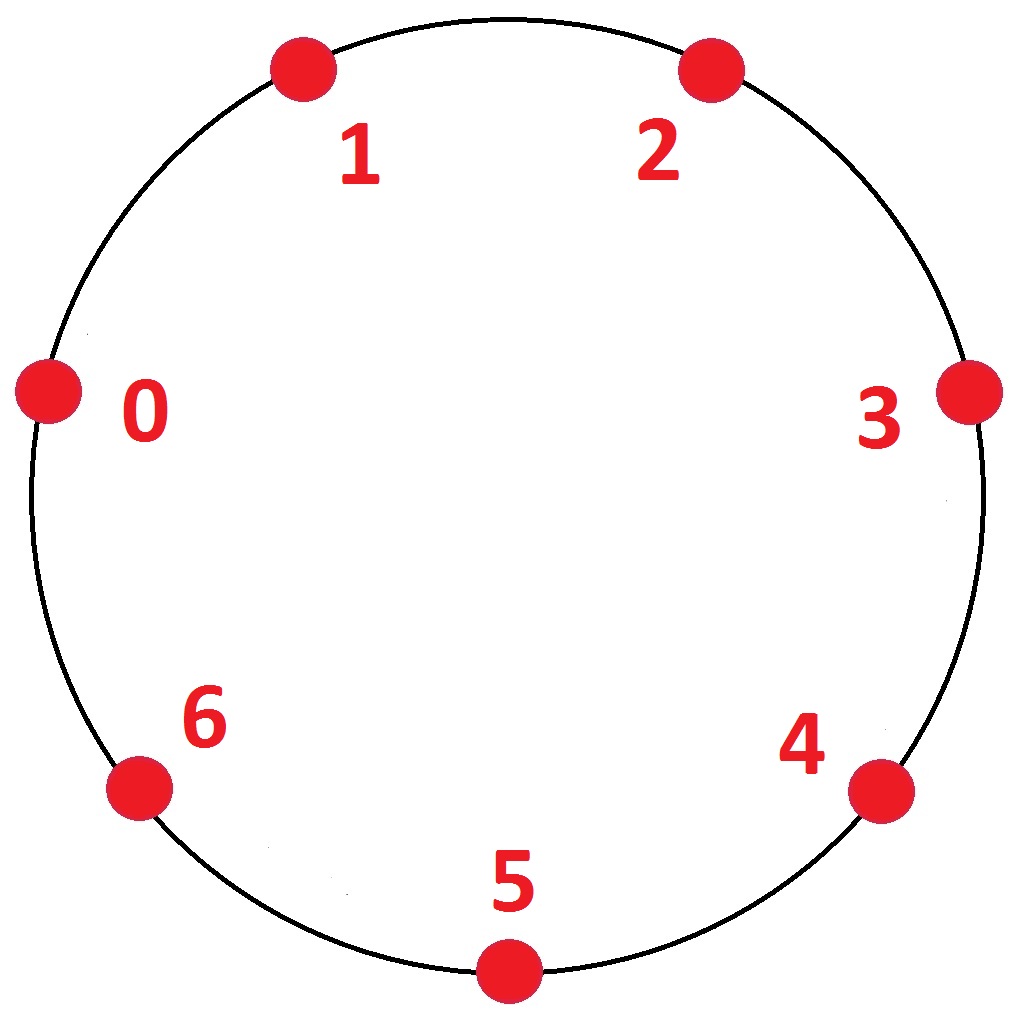}\caption{{A simple graph with $7$ vertices
which approximates a circular one-dimensional space.}}
\label{cerchio}
\end{figure}

\noindent In the discretized theory, the integral over points becomes a sum over
vertices of the graph. Similarly, the derivative becomes a finite difference.
For simplicity, we start with a one-dimensional graph: it's easy to see how the
derivative operator is proportional to an antisymmetric matrix $\tilde{M}$ whose
elements are different from zero only immediately above the diagonal (where they count +1),
and immediately below (where they count -1). We can see this, for example, in a
\lq\lq toy-graph'' formed by only $7$ separated vertices (figure \ref{cerchio}).
The argument remains true while increasing the number of vertices.

\footnotesize
\ba
\partial\varphi (p_i) &=& \fr 1 {2\Delta} \left(
\begin{array}
[c]{ccccccc}%
0 & +1 & 0 & 0 & 0 & 0 & -1 \\
-1 & 0 & +1 & 0 & 0 & 0 & 0 \\
0 & -1 & 0 & +1 & 0 & 0 & 0 \\
0 & 0 & -1 & 0 & +1 & 0 & 0 \\
0 & 0 & 0 & -1 & 0 & +1 & 0 \\
0 & 0 & 0 & 0 & -1 & 0 & +1 \\
+1 & 0 & 0 & 0 & 0 & -1 & 0 \\
\end{array}
\right)  \left(
\begin{array}
[c]{c}%
\varphi\left(  0\right) \\
\varphi\left(  1\right) \\
\varphi\left(  2\right) \\
\varphi\left(  3\right) \\
\varphi\left(  4\right) \\
\varphi\left(  5\right) \\
\varphi\left(  6\right) \\
\varphi\left(  7\right)
\end{array}
\right) = \fr 1 {2\Delta}\left(
\begin{array}
[c]{c}%
\varphi\left(  1\right)  -\varphi\left(  6\right) \\
\varphi\left(  2\right)  -\varphi\left(  0\right) \\
\varphi\left(  3\right)  -\varphi\left(  1\right) \\
\varphi\left(  4\right)  -\varphi\left(  2\right) \\
\varphi\left(  5\right)  -\varphi\left(  3\right) \\
\varphi\left(  6\right)  -\varphi\left(  6\right) \\
\varphi\left(  0\right)  -\varphi\left(  5\right)
\end{array}
\right)\nonumber\\
&& \label{forma} \ea

\normalsize

\noindent $\Delta$ is the length of graph edges. In the continuous limit, $\Delta \ra 0$
(where matricial product turns into a convolution), we obtain

\ba
\pa \varphi (x) &=& \lim_{\Delta \ra 0}\fr 1 {2\Delta} \int \tilde{M}(x,y) \varphi (y) dy \nonumber \\
\pa \varphi (x) &=& \lim_{\Delta \ra 0}\fr 1 {2\Delta} \int \left[ \delta (y-(x+\Delta)) - \delta (y-(x-\Delta))\right] \varphi (y) dy \nonumber \\
\pa \varphi (x) &=& \lim_{\Delta \ra 0}\fr {\varphi(x+\Delta)-\varphi(x-\Delta)} {2\Delta} = \pa\varphi (x)
\ea
In this way our definition is consistent with the usual definition of derivative.

While increasing the number of points, a $(-1)$ still remains in the up right corner
of the matrix, and a $(+1)$ in the down left corner as well.
To remove those two non-null terms, it is sufficient to make them unnecessary,
by imposing boundary conditions that make the field null in the first and in the last point.

In fact we can describe an open universe (a straight line in one dimension),
starting from a closed universe (a circle) and making the radius to tend to infinity.
Hence we see that the conditions of null field in the first and in the last point become
the traditional boundary conditions for the Standard Model fields.

\begin{notation}
Note that in spaces with more than one dimension, a derivative matrix $\tilde{M}_\mu$
assumes the form (\ref{forma}) only if we number the vertices progressively
along the coordinate $\mu$. However, two different numberings can be always related
by a vertices permutation.
\end{notation}

\section{Local translations \label{ord}}

In this section we introduce a non-local gauge field for local
translations. Moreover we see that such field is enough
to describe a covariant quantum theory (ie a quantum theory of
gravity).

A global translation for a scalar field $\varphi$ is given by

$$\varphi'(x) = \varphi(x+\l) = e^{\l^\mu \pa_\mu} \varphi(x).$$

\noindent A local translation is then given by

$$\varphi'(x) = \varphi(x+\l(x)) = E(\l(x)) \varphi(x)$$

\noindent where

$$E(\l) = 1 +\l^\mu \pa_\mu + \fr 1 {2!} \l^{\mu_1}\l^{\mu_2} \pa_{\mu_1} \pa_{\mu_2} +\ldots + \fr 1{n!} \l^{\mu_1}\ldots \l^{\mu_n} \pa_{\mu_1} \ldots \pa_{\mu_n}$$

\noindent While for global translations we have $(\pa_\nu \varphi)'(x) = (\pa_\nu \varphi)(x+\l)= e^{\l^\mu \pa_\mu} \pa_\nu \varphi (x)$, for local translations we have $\pa_\nu \varphi'(x) = \pa_\nu \varphi (x+\l(x)) \neq E(\l) \pa_\nu \varphi (x)$. Conversely:

$$\pa_\nu \varphi' (x) = E(\l(x))\pa_\nu \varphi (x) + \left[ \pa_\nu, E(\l(x))\right] \varphi(x)$$

\noindent We take $\pa' = \pa$ and $(\pa_\nu \f)' = \pa_\nu \f'$ 	
insofar as translations are considered transformations of fields and not of coordinates. Accepted this we can define a covariant derivative

$$D_\nu = \pa_\nu + G_\nu$$
where $G_\nu$ is a gauge field with transformation law

$$G_\nu \ra G'_\nu = E(\l) G_\nu E^{-1}(\l) - \left[\pa_\nu, E(\l)\right]E^{-1}(\l).$$
The transformation law for $D$ is easily calculated:

\ba (D\varphi)' &=& D'_\nu \f' = (\pa'_\nu + G'_\nu)\f' = (\pa_\nu + G'_\nu)\f' = \nonumber \\
    &=& E(\l)(\pa_\nu \f) + \left[\pa_\nu, E(\l) \right] \f + E(\l)G_\nu \f - \left[ \pa_\nu, E(\l)\right] \f = \nonumber \\
    &=& E(\l) (\pa_\nu + G_\nu)\f = E(\l) (D_\nu \f) \ea
Hence

$$D'_\nu = E(\l)D_\nu E^{-1}(\l) = D_\nu$$

\noindent What is $E^{-1}(\l)$? Consider that

$$\f (x) = E^{-1}(\l) \f'(x) = E^{-1}(\l) \f(x+\l(x))$$

\noindent Defining a new coordinate $x' = x +\l(x)$ we obtain

$$\f (x'-\l(x(x'))) = E^{-1}(\l) \f(x')$$
\ba && E^{-1}(\l) = E(-\l') \nonumber \\
     && \quad = 1- \l^\mu \fr \pa {\pa x'^\mu} + \fr 1 {2!} \l^{\mu_1} \l^{\mu_2} \fr \pa {\pa x'^{\mu_1}} \fr \pa {\pa x'^{\mu_2}} + \ldots + \fr {(-1)^n} {n!} \l^{\mu_1} \ldots \l^{\mu_n} \fr \pa {\pa x'^{\mu_1}} \ldots \fr \pa {\pa x'^{\mu_n}} \nonumber \ea

$$\fr \pa {\pa x'^\mu} = \left. \left( \fr {\pa(x + \l(x))} {\pa x} \right)^{-1} \right.^\nu_{\pt \mu} \pa_\nu$$

\noindent What's about $\f'(x')$ ?

\ba \f' (x') &=& \f'(x+\l(x)) = \f (x' + \l(x')) = \nonumber \\
        &=& \f (x + \l(x) + \l(x + \l(x))) = \f (x + \Lambda (x)) = \nonumber \\
        &=& E(\Lambda(x)) \f(x) \ea
with $\Lambda(x) = \l(x) + \l(x+\l(x))$.

Take now a field $G_\nu$ which is pure gauge, explicitly $G_\nu = 0$. Under
a translation $x \ra x + \Lambda(x)$ we have

\ba D_\nu &=& \pa_\nu \nonumber \\
    G'_\nu &=& -\left[ \pa_\nu, E(\Lambda)\right]E(-\Lambda') \nonumber \\
    D'_\nu &=& E(\Lambda)\pa_\nu E(-\Lambda') \nonumber \\
    &=& \left. \left( \fr {\pa (x+\l(x))} {\pa x} \right)^{-1}\right.^{\mu}_{\pt \nu} \pa_\mu = \fr {\pa}{\pa(x+\l(x))^\nu} = \fr \pa{\pa x'^\nu} \ea

\noindent The last relation is been obtained in three steps:

$$\pa_\nu E(-\Lambda')\underbrace{E(\Lambda)\f(x)}_{\f'(x')} = \pa_\nu \f(x)$$
$$E(\Lambda) \pa_\nu E(-\Lambda')\underbrace{E(\Lambda)\f(x)}_{\f'(x')} = E(\Lambda) \pa_\nu \f(x)$$
$$D'_\nu \f'(x') = (\pa_\nu \f)(x+\Lambda(x)) = (\pa_\nu \f)'(x+\l(x)) = \fr \pa{\pa x'^\nu} \f'(x')$$

\noindent Note that we have recovered the usual definition of $\pa'$, ie

$$\pa'_\nu = \fr \pa {\pa x'^\nu} = \fr \pa {\pa(x+\l(x))^\nu}$$

\noindent We see that the action of a local translation $x \ra x + \Lambda(x)$
corresponds to the diffeomorphism $x \ra x' = x + \l(x)$ with
$\Lambda(x) = \l(x) + \l(x+\l(x))$. Hence we have two choices
to obtain invariance under diffeomorphisms:

\begin{enumerate}
\item To define $e^\mu_a (x)$ with $e'^\mu_a (x') = \fr {\pa x'^\mu}{\pa x^\nu} e^\nu_a (x)$, in such a way to have

$$\mathfrak{S} = \int d^4 x\, \eta^{ab} e^\mu_a e^\nu_b (\pa_\mu \f)(\pa_\nu \f)$$
invariant;

\item To introduce a vector field $G_\nu$ which transforms as an ordinary gauge field

$$G_\nu \ra G'_\nu = E(\Lambda) G_\nu E(-\Lambda') - \left[\pa_\nu, E(\Lambda)\right]E(-\Lambda')$$
in such a way to have

$$\mathfrak{S} = \int d^4x \,\eta^{\mu\nu} (D_\mu \f) (D_\nu \f)$$
invariant.
\end{enumerate}
The last case requires a bit of attention. In fact $\mathfrak{S}$ is invariant if and only if the field $\f$ transforms according to the rule

$$\f(x) \ra \f'(x') = E(\Lambda) \f(x) E(-\Lambda')$$
instead of

$$\f(x) \ra \f'(x') = E(\Lambda) \f(x).$$
However $\f'(x') = E(\Lambda) \f(x)$ in both cases when it applies to a constant field.
Substituting in $\mathfrak{S}$ we find:

\ba \mathfrak{S}' &=& \int d^4 x\,\eta^{\mu\nu} E(\Lambda)(D_\mu \f)\underbrace{E(-\Lambda')E(\Lambda)}_{=1} (D_\nu \f)E(-\Lambda') \nonumber \\ &=& \int d^4 x\,\eta^{\mu\nu} E(\Lambda)(D_\mu \f) (D_\nu \f)E(-\Lambda')\nonumber \ea

\noindent In the discrete framework the operators $E$ are simple matrices and $\int d^4 x$ acts as a trace. Hence we can apply the cyclicity property:

\ba \mathfrak{S}' &=& \int d^4 x\,\eta^{\mu\nu} \underbrace{E(-\Lambda') E(\Lambda)}_{=1}(D_\mu \f) (D_\nu \f)\nonumber \\
 &=& \int d^4 x\,\eta^{\mu\nu} (D_\mu \f) (D_\nu \f) = \mathfrak{S} \nonumber \ea

\noindent Pay attention that $(\pa_\nu \f)$ is equal to $\pa_\nu \f - \f \pa_\nu$ and so it corresponds to $[\tilde{M}, \f]$ with $\f$ in diagonal representation. At the same time the appearance of $E(\Lambda)$ is very simple: it has a $1$ for every $p_i$ in the crossing between the row $p_i$ and the column $p_i + \Lambda (p_i)$.
Other entries are zero. Finally $E(-\Lambda')$ is calculated as the inverse of $E(\Lambda)$.

In presence of fields with spin we have to add the usual spin connection $\w$
in order to compensate local Lorentz transformations. It is necessary because diffeomorphisms
cause simultaneously a local shift of coordinates and a local \lq\lq rotation'' of axes.
Clearly this last doesn't affect scalar fields.
Putting all together:

$$D_\mu = \pa_\mu + G_\mu + \w_\mu^{ij} \qquad\quad i,j=0,1,2,3 \qquad \w^{ij} = - \w^{ji}$$

\noindent Obviously all fields transform homogeneously under local translations:

\ba \overline{\psi} &\ra& \overline{\psi}' = E \overline{\psi} E^{-1} \nonumber \\
    \psi &\ra& \psi' = E \psi E^{-1} \nonumber \\
    A_\mu &\ra& A'_\mu = E A_\mu E^{-1} \nonumber \\
    D_\mu \f &\ra& (D_\mu \f)' = E (D_\mu \f) E^{-1} \ea

\noindent We see that all fields can be local only in a specified gauge. In other words, given a local
field $\f$ in the diagonal representation, its transformed $\f' = E\f E^{-1}$ is highly non local.

\section{Ricci scalar and Feynman diagrams}

\begin{figure}[ptbh]
\centering\includegraphics[width=1\textwidth ]{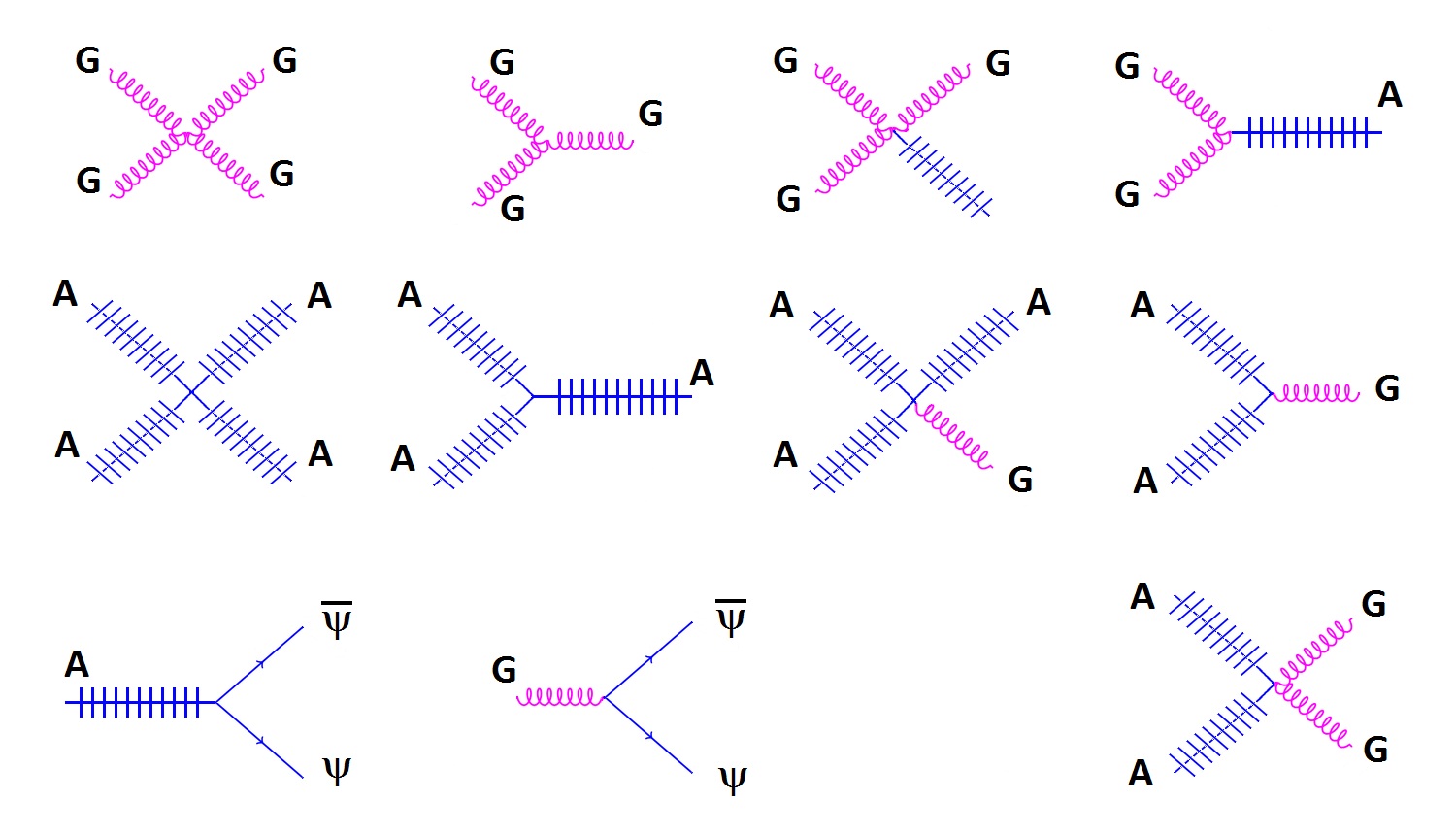}\caption{Feynman diagrams in Quantum Gravity.}
\label{diagrams}
\end{figure}

In some gauge the Ricci Scalar is given by

$$R = \a \left[(\pa_\rho + A_\rho), (\pa_\s + A_\s)\right] \left[G_\mu, G_\nu\right]\eta^{\mu\rho}\eta^{\nu \s} \qquad\quad \a \in \mathbf{R}$$
However, the covariant combination

$$K = \a [D_\mu, D_\nu][D_\rho, D_\s] \eta^{\mu\rho}\eta^{\nu \s}$$
is equivalent to Ricci scalar plus a topological term (usually called \lq\lq Gauss-Bonnet term'').
Moreover we can extract from $K$ the propagators of $G$ and $A$ as it happens in
ordinary gauge theories. Adding the expected fermionic term $\bar{\psi}\g^\mu D_\mu \psi$,
the Feynman vertices of quantum gravity are the ones in figure \ref{diagrams}.

The price for treating Quantum Gravity as a gauge theory is the inclusion of non-local fields.
Explicitly, the path integral must contain not only integrations over diagonal fields
$\f(x)$, $\psi(x)$ $\bar{\psi}(x)$, $A_\mu(x)$ and $G_\mu (x)$, but also over non-local fields
$\f(x,y)$, $\psi(x,y)$ $\bar{\psi}(x,y)$, $A_\mu (x,y)$ and $G_\mu (x,y)$.

In this way we recover the fundamental result of \textbf{Arrangement Field Theory}, ie that \emph{metric appears when locality principle is imposed}. Fortunately we can choose a local gauge, which in contrast gives non-local ghost fields.

\section{Structure constants}

Our target is now the calculation of structure constants for the group of local translations.
We start from a gauge field $G_\mu$ which is pure gauge:

$$G^{(\a)}_\mu = -[\pa_\mu, E(\a)]E(-\a') = -\pa_\mu + E(\a)\pa_\mu E(-\a')$$

\noindent Applying it to $\f(x')$ with $x' = x +\a(x)$:

\ba G^{(\a)}_\mu \f(x') &=& -\pa_\mu \f(x') + E(\a)\pa_\mu \f(x) \nonumber \\
                        &=& -\fr {\pa x'^\nu}{\pa x^\mu} \fr \pa {\pa x'^\nu} \f(x') + \fr \pa {\pa x'^\mu} \f(x') \nonumber \\
                        &=& -\fr \pa {\pa x'^\mu} \f(x') - (\pa_\mu \a^\nu (x))\fr \pa {\pa x'^\nu} \f(x') + \fr \pa {\pa x'^\mu} \f(x') \nonumber \\
                        &=& - (\pa_\mu \a^\nu (x))\fr \pa {\pa x'^\nu} \f(x') \\
                        &\pt& \nonumber \\
\Longrightarrow G^{(\a)}_\mu &=& - \fr{(\pa_\mu \a^\nu )} {\left[ \d^\nu_\g + \pa_\g \a^\nu \right]} \pa_\g \nonumber \ea

\noindent We can write

\be -\fr{(\pa_\mu \a^\nu )} {\left[ \d^\nu_\g + \pa_\g \a^\nu \right]} \overset{!}{=} \tilde{\a}^i_\mu f^\g_i (x) \label{form} \ee

\noindent where we have introduced a base of functions $\{f^\g_i (x)\}$ for a suitable Hilbert space and complex constants $\tilde{\a}^i_\mu$. We conclude that a generic $G^{(\a)}_\mu$ has form $\tilde{\a}^i_\mu f^\g_i (x)$ with
$\tilde{\a}^i_\mu$ not restricted by (\ref{form}). It follows:

\ba [ G^{(\a)}_\mu, G^{(\b)}_\nu ] &=& \tilde{\a}^i_\mu \tilde{\b}^j_\nu (f^\s_i \pa_\s f^\g_j - f^\s_j \pa_\s f^\g_i) \pa_\g \nonumber \\
   &=& \tilde{\a}^i_\mu \tilde{\b}^j_\nu \Gamma^l_{\pt ij} f^\g_l \pa_\g \ea

\noindent where $\Gamma^l_{\pt ij}$ are structure constants of local translations with

$$\Gamma^l_{\pt ij} = \Big( f^\s_i \pa_\s f^\g_j - f^\s_j \pa_\s f^\g_i \,\,\, , \,\,\, f^\g_l \Big).$$

\noindent Summarizing, local translations define an infinite dimensional non-abelian group with generators $f^\g_i \pa_\g$ and structure constants $\Gamma^l_{\pt ij}$.

Consider now the commutator between $G^{(\a)}_\mu$ and the spin-connection $\w^{ab}_\nu (x) = \w^{abj}_{\nu\t} f^\t_j (x)$:

$$\left[ G^{(\a)}_\mu, \w^{ab}_\nu \right] = \tilde{\a}^i_\mu \w^{abj}_{\nu\t} E^{l}_{\pt ij} f^\t_l $$
$$E^{l}_{\pt ij} = \Big( f^\s_i \pa_\s f^\tau_j \,\,\, , \,\,\, f^\tau_l \Big).$$

\noindent We understand that local Poincar\'e doesn't factorize in Lorentz $\times$ translations, due to non-zero
structure constants $E^{l}_{\pt ij}$. Lorentz and translations remain subgroups, but they are no longer independent.

We choose as base functions $f_i (x) = \fr 1 {(2\pi)^2}\l^{(\t(i))} e^{ip(i)_{\nu} x^\nu}$, where $\l^{(\t)}$ is a column array with $1$ in the $\t$-th entry and $0$ elsewhere $(\l^{(\t)\s} = \d^{\t\s})$. Moreover we impose $( \l^\s, \l^\g ) = \d^{\s\g}$. The structure constants result

\ba \Gamma^l_{\pt ij} &=& \int d^4x\, \fr {1} {(2\pi)^6} \left[ \l^{(\t (i))\s} e^{ip(i)_{\nu}x^\nu} (ip(j)_{\s}) e^{ip(j)_{\nu}x^\nu} (\l^{\t(j)}, \l^{\t(l)}) \right. -  \nonumber \\
&&\qquad\qquad\quad - \left. \l^{(\t (j))\s} e^{ip(j)_{\nu}x^\nu} (ip(i)_{\s}) e^{ip(i)_{\nu}x^\nu} (\l^{\t(i)}, \l^{\t(l)}) \right]  e^{-ip(l)_{\nu} x^\nu} \nonumber \\
      &=& \int d^4x\, \fr {i} {(2\pi)^6} (\d^{\t(i)\t(l)} p(j)_{\t(i)}- \d^{\t(j)\t(l)} p(i)_{\t(j)}) e^{i(p(i)_{\nu}+ p(j)_{\nu} - p(l)_{\nu}) x^\nu} \nonumber \\
      &=& \fr {i} {(2\pi)^2} \Big(\d^{\t(i)\t(l)} p(j)_{\t(i)}- \d^{\t(j)\t(l)} p(i)_{\t(j)}\Big) \d^4 (p(i)+p(j)-p(l)) \nonumber \ea
      
\noindent In the two last lines there is no sum over repeated indices $\t(i)$ and $\t(j)$. The $\d^4$ implies the usual momenta conservation in trivalent vertices. More important is to note that $\Gamma \sim p$, while in usual gauge theories we have $\Gamma \sim 1$. Fortunately this defect is compensated by propagator. In fact

$$\text{Kinetic term} \sim (\pa G)^2 \sim (\pa \tilde{\a} \pa)^2 \sim (p^2 \tilde{\a})^2 \sim p^4 \tilde{\a}^2$$
$$\Rightarrow \text{Propagator} \sim \fr 1 {p^2} \cdot \fr 1 {p^2}$$
and so the extra $\fr 1 {p^2}$ compensates the extra $p$ in structure constants.

\section{Conclusion}

At this point we have a quantum theory of gravity which resembles ordinary gauge theories. This implies that quantum gravity has the same superficial degree of divergence of gauge theories and then it is apparently quantizable. You just have to try to calculate the first gravitational amplitudes.

\end{document}